\documentclass[preprint]{aastex}
\tighten
\input{epsf.sty}

\newcommand{\solar}{\ifmmode_{\mathord\odot}\else$_{\mathord\odot}$\fi} 
\newcommand{\kms}{km\thinspace s$^{-1}$}     
\newcommand{\degre}{\ifmmode^\circ\else$^\circ$\fi}  
\newcommand{\arcs}{\ifmmode {'' }\else $'' $\fi}  
\newcommand{\arcm}{\ifmmode {' }\else $' $\fi}    
\newcommand{\mper}{\ifmmode \buildrel m\over . \else $\buildrel m\over .$\fi} 
\newcommand{\hper}{\ifmmode \rlap.^{h}\else $\rlap{.}^h $\fi} 
\newcommand{\sper}{\ifmmode \rlap.^{s}\else $\rlap{.}^s $\fi} 
\newcommand{\arcsper}{\ifmmode \rlap.{'' }\else $\rlap{.}'' $\fi} 
\newcommand{\arcmper}{\ifmmode \rlap.{' }\else $\rlap{.}' $\fi} 
\newcommand{\bull}{\vrule height .9ex width .8ex depth -.1ex}
\newcommand{\et}{{\it et\thinspace al.}}   

\shortauthors{Rosenberg and Schneider}
\shorttitle{Properties of HI-Selected Galaxies}

\begin{document}

\title{The Arecibo Dual-Beam Survey: Arecibo\footnote{The Arecibo Observatory
is part of the National Astronomy and Ionosphere Center, which is operated by
Cornell University under cooperative agreement with the National Science
Foundation. in Puerto Rico.} \ and VLA\footnote{The National Radio
Astronomy Observatory is a facility of the National Science Foundation operated
under cooperative agreement by Associated Universities, Inc.} \ Observations }
 
\author{Jessica L. Rosenberg \& Stephen E. Schneider}
\affil{Department of Astronomy, University of Massachusetts, Amherst, MA 01003}
\email{rosenber@umbriel.astro.umass.edu}
\email{schneide@messier.astro.umass.edu}

\begin{abstract}

The Arecibo Dual-Beam Survey is a ``blind" 21 cm search for galaxies covering
$\sim$430 deg$^2$ of sky. We present the data from the detection survey as
well as from the follow-up observations to confirm detections and improve
positions and flux measurements. We find 265 galaxies, many of which are
extremely low surface brightness. Some of these previously uncataloged galaxies
lie within the zone of avoidance where they are obscured by the gas and dust in
our Galaxy. 81 of these sources are not previously cataloged optically
and there are 11 galaxies that have no associated optical counterpart or are
only tentatively associated with faint wisps of nebulosity on the Digitized Sky
Survey images. We discuss the properties of the survey and in particular we
make direct determinations of the completeness and reliability of the sample.
The behavior of the completeness and its dependencies is essential for
determining the HI mass function. We leave the discussion of the mass function
for a later paper, but do note that we find many low surface brightness galaxies
and 7 sources with M$_{HI}$ $<$ 10$^8$ M\solar. 

\end{abstract}

\section{INTRODUCTION}

 \par We present the Arecibo Dual-Beam Survey (ADBS), a ``blind" 21cm survey for
galaxies. ``Blind" HI surveys provide a look at galaxy populations that is
unbiased by stellar populations. 
Many previous searches have tried to determine whether there is a hidden
population of HI-rich galaxies. They have looked in voids (Szomoru \et\ 1996,
Hulsbosch 1987, Krumm \& Brosch 1984) in clusters (Banks \et\ 1999, Szomoru \et\
1994, Weinberg \et\ 1991), in groups (Kraan-Korteweg \et\ 1999, Haynes \&
Roberts 1979, and Lo \& Sargent 1979) and at random directions in the sky
(Kilborn \et\ 1999, Spitzak \& Schneider 1998, Zwaan \et\ 1997, Henning 1992,
Fisher \& Tully 1981). These surveys have not been able to convincingly
determine the nature of the population of low HI-mass galaxies. Zwaan \et\
(1997) and Spitzak \& Schneider (1998) undertook ``blind" HI surveys, similar to
the one discussed here, but neither had enough low mass sources to make a strong
claim about the shape of the mass function. Schneider \et\ (1998) combined the
two data sets and found evidence for a rise in the number of low mass sources,
but the results are not definitive. The HIPASS survey, with its coverage of the
entire Southern sky, should also contribute to our understanding of the HI mass
function. The most recently published mass function contains 263 galaxies,
comparable to the number in this survey (Kilborn \et\ 1999) but only has 2
galaxies with M$_{HI} < $ 10$^8$ M\solar\ (converting to H$_0$ = 75 \kms\
Mpc$^{-1}$, which is used throughout this paper). Although the debate about the
number of low HI mass sources persists, these galaxies certainly do exist. We
detect 7 sources with HI masses $<$ 10$^8$ M\solar, almost as many as all of the
previous ``blind" HI surveys combined.

\par The ADBS is one of the largest and most sensitive surveys to date.  
Using the Arecibo telescope we were able to achieve an rms of 3-4 mJy at a 
resolution of 32 \kms\ in only 7 seconds and covered $\sim$430 deg$^2$. The
ADBS covers substantially more volume than the deeper Spitzak \& Schneider
(1998) and  the Zwaan \et\ (1997) surveys, and is much more sensitive than the
larger Parkes  HIPASS survey (Kilborn \et\ 1999). HIPASS will cover the entire
southern sky when  it is completed, but it has a sensitivity of only 13 mJy at a
resolution of 13.2 \kms\ and has trouble with the confusion of multiple HI
sources within its 14\arcm\ beam. With the low sensitivity,
the Parkes survey is only able to detect low mass sources at very low redshifts
where distances are difficult to determine.

\par In this paper we will discuss the HI data and our efforts to quantify the
completeness and reliability of this sample. This is the first survey to use
synthetic sources to derive an empirical relation between the flux, line width, 
and
completeness of the survey. This will allow us to determine an HI mass function
even if we do not understand all of the circumstances under which sources are
not detected. The HI mass function will be discussed in Rosenberg \& Schneider 
(in preparation; hereafter paper II).

We also present the optical identifications of the ADBS sources from the
Digitized Sky Survey (DSS) images. We identify 81 HI-rich galaxies that are not
in optical catalogs. Out of these 81 sources, 11 are not identified because they
are heavily obscured by the Galactic plane. An additional 11 sources are at high
Galactic latitude where there is no clear optical identification. In most of
these cases, there is a hint of nebulosity that might be associated, a
possible background source, or a bright star, but no convincing association.
Even if these optical counterparts are associated, they are extremely faint
relative to the mass of HI present. These 11 sources represent the most extreme
cases in the survey, but we detect many more sources that were previously
uncataloged because they are very faint.

\par We describe the survey in \S 2 and the VLA and Arecibo follow-up
in \S 3. The data and data tables are in \S 4 followed by a discussion
of the completeness in \S 5 and the properties of the sample in \S 6.
The results of the survey are summarized in \S 7. 

\section{OBSERVATIONS}

\subsection{The Detection Survey}

\par The goals of our survey are (1) to determine whether there are
classes of galaxies that have been overlooked previously and (2) to help tie
down the HI mass function with the larger number of sources in this sample. To
achieve these goals a ``blind" HI survey needs to detect galaxies down to the 
lowest
possible HI fluxes. At least equally important is a detailed understanding of
the survey limitations. The statistics of previous surveys have been
insufficient to definitively establish the shape of the HI mass function at both
the high and the low mass ends. To improve the statistics one needs to cover a
large volume of space. For a fixed total observing time more angular 
coverage means less sensitivity but greater volume coverage for a given HI mass
since area
$\propto$ t$^{-1}$ but depth $\propto$ t$^{1/4}$ where t is the
integration time per point. Therefore the effective volume covered in a fixed 
observing time $\propto$ t$^{-1/4}$. This might suggest that integration times 
should be made as short as possible, except that as sensitivity drops, low mass
sources are detectable at smaller redshifts where the distance uncertainty
becomes more significant and confusion with Galactic HI becomes a problem. 

In trying to detect sources there is also a trade off between completeness and
reliability. As digging deeper into the noise improves completeness, but causes
the selection of more and more false sources. Likewise, eliminating narrow
features that appear to be interference will eliminate some narrow linewidth
sources. A  balance must be achieved between the number of sources that can be
verified and going deep enough to detect some of the fainter sources. In the
end, the importance is in quantifying these selection effects. We discuss our
completeness  and reliability in \S 5.

\par There are many factors that go into determining which telescope is most
suitable for a given project including collecting area, beam size, and system
temperature. The sensitivity is proportional to the collecting area and
inversely proportional to the system temperature. The collecting area is also
inversely proportional to the beam size, so the decrease in beam size for a
large telescope increases the time required to  cover a given area of the sky.
If the system temperatures are the same, telescopes of different sizes can cover
the same area with the same sensitivity in equal times, however if the beam
size if too large then source confusion becomes a problem. Confusion may be
resolved later, but then it may be difficult to determine whether the galaxies
would have been detected independently. Fortunately, at the typical distance of 
sources detected in only 7 seconds of integration, Arecibo's 3.3\arcm\ beamsize
is usually able to distinguish neighboring galaxies. 

During these observations, Arecibo was being upgraded and had limited mobility. 
This was an advantage for this project because it made large amounts of
telescope time available for driftscan searches. Observing in driftscan
mode is an efficient method when the goal is to cover a large area on the sky 
and Arecibo is sufficiently sensitive that useful depths can be reached in the
time a source takes to drift through the telescope beam.
The lack of tracking also meant that two feeds could be used simultaneously. We 
used the 21 cm and 22 cm circular polarization feeds,
which have now been replaced by the Gregorian reflector. These were located on 
opposite
corners of the carriage house and pointed 1.6\degre\ apart on the sky. Our
ability to use two feeds doubled the area surveyed, although it did require 
coarsening the velocity resolution to twice that in the Slice Survey (Spitzak 
\& Schneider 1998).

\par One of the major challenges of any single-dish HI survey is to distinguish
between real sources and radio interference. We used three tactics to address
this problem while also increasing the volume coverage and improving the
signal-to-noise: (1) Comparing data from the 21 and 22 cm feeds taken
simultaneously -- the interference usually enters through far sidelobes and
appears in both feeds while real sources appear in only one. (2) Comparing left
and right circular polarizations in each feed -- the average of the two
polarizations provides improved signal-to-noise while the difference is an
additional check for interference since it is often highly polarized. (3)
Observing each declination strip twice, on separate days, providing confirmation
of the source detections.

\par While the driftscan technique allowed us to cover a large volume of space
rapidly, it also gave us uncertainties of up to $\sim$7$\arcm$ in the declination
position because the sources could be in the main beam or first sidelobe (or
rarely farther out if the source was very bright at 21 cm).
The lack of declination information meant that the sources needed to be
reobserved to determine their positions and fluxes. We subsequently
followed-up the identifications at the Very Large Array (VLA) and at Arecibo
(\S 3). 

\par The Arecibo 305m telescope was dedicated to the ADBS for approximately 500
hours between December 1993 and February 1994. These data consist of
$\sim$300,000 spectra taken every 7 seconds and cover nearly 24 hours of time
in each of 30 declination strips. The total sky coverage, shown in Figure 1, is
approximately 430 deg$^2$ in the 3.3\arcm\ main beam. In this figure each
strip is surrounded by two thin lines showing the positions of the sidelobes.
The thick lines denote the regions covered twice while the thin lines denote the
regions of single coverage. The actual sky coverage is the convolution of the
sky coverage in Figure 1 with the beam shape. Figure 2 shows this convolution
for the beginning of a driftscan. The contours indicate the total integrated
detection sensitivity at each position, normalized to unity along the beam's
center-line.

\par The velocity coverage for the survey was --654 to 7977 \kms. This coverage
was achieved over 512 channels in each of two polarizations for each of two
feeds, totalling 2048 channels. The channel spacing was 16.9 \kms\ and the
resolution was 33.8 \kms\ after Hanning smoothing.

\subsection{The ADBS Data Reduction}

\par The data reduction for ADBS spectra was a several step process which was
run as an external FORTRAN module to ANALYZ, Arecibo Observatory's data
reduction package. The data are
driftscans and the adjacent spectra behave as off scans for each spectrum. The
resulting noise, after the data have been reduced, is 3-4 mJy. We review the
basic steps taken in the data reduction here: 
\par
\noindent (1) The total power in the spectra was adjusted according to the
recorded power counters. 
\par
\noindent (2) An average ``off-source" spectrum was created from all of the
spectra in a scan (usually 300) after eliminating any outliers--either due to
interference signals in individual channels or continuum flux across all 
channels.    
\par
\noindent (3) Each spectrum had the ``off" subtracted and then was normalized by
it: $(on-off)/off$.
\par 
\noindent (4) Synthetic sources were added so that they would go through the
flattening procedure. The synthetic sources are discussed further in \S 5.
\par
\noindent (5) The continuum emission was estimated from the total power counters
for each spectrum and then subtracted using the bandpass shape of each feed. The
bandpass is assumed to be Gaussian centered at 1408.5 and 1398.5 MHz
respectively, with a half-power width of 52 MHz for both feeds. These values
were tested and adjusted until subtraction gave a flat response.
\par 
\noindent (6) After excluding channels with strong signals compared to 
surrounding channels, a linear fit was made to each spectrum and subtracted.
\par
\noindent (7) Two ``off" scans were created from data on either side of each
spectrum. These ``offs" consisted of 19 spectra separated from the ``on"
spectrum by 6 spectra. If either of the ``offs" produced an improvement they
were used. The idea was to remove some of the longer time-scale problems
(standing waves and interference), that are matched in nearby spectra without
removing the source. 
\par
\noindent (8) The spectra were Hanning smoothed.
\par 
\noindent (9) At this point the baseline looked very flat in most cases but
there were a few exceptions where there was still low level waviness to the
baselines. A final polynomial fit was designed to eliminate these underlying
variations while ignoring any regions where there were rapid variations. A
fifth-order polynomial was fit to each spectrum excluding any regions where
there was a change of 7 mJy ($\sim2\sigma$) or greater over a range of
$\sim$32\kms. Tests showed that this yielded a highly constrained fit that
ignored weak, narrow signals, although it could reduce the apparent flux from
wide, weak signals, especially when they were near a bandpass edge.

\par
\noindent (10) Images were Hanning smoothed in right ascension.

\par The ADBS survey was analyzed both by-eye and using the SExtractor (Bertin
\& Arnouts 1996) computer algorithm. In order to facilitate both of these
analyses, groups of 300 spectra were turned into two-dimensional grayscale FITS
images. Figure 3 is an example of one of these images. The images are 601 rows
high with 300 spectra from the 21 cm feed in the top half of the image and 300
22 cm feed spectra, taken at the same time as the first 300, in the bottom half.
The topmost row of the image (row 601) gives the center velocity of each column
or correlator segment. The columns are broken into two sections; the first 512
columns represent the 512 correlator channels averaged over the 2 polarizations,
the last 512 columns contain the absolute value of the difference between the
polarizations used as an interference check. There are some extra columns
between the sum and difference of the polarizations which contain information
about the baseline fits for each spectrum, the strength of the continuum, and
the spectrum coordinates so that the information is easily available while
searching for galaxies. 

\par There are a few features to note in Figure 3. The first is that the
prominent black and white ``candy cane" near the left edge of the image is the
Milky Way. The Milky Way emission does not subtract completely from the
difference between the two polarizations (located just to the right of the image
center) probably because the response to such strong emission is not always
linear. The vertical stripes through the averaged polarizations, which stand out
better in the polarization difference, are due to continuous interference which
is variable in strength.  The leftmost interference stripe shows up in the
polarization difference, but not in the averaged polarizations because it is
weak and fairly constant in strength. Momentary interference appears as two
blips, one in the 21cm feed and the other in the 22 cm feed. These blips are
also apparent in the polarization difference image. Other images occasionally
show a wavy background due to standing wave patterns set up in the telescope. In
these regions, galaxy detection is difficult but the measured RMS noise is also
correspondingly bigger. On rare occasions, there is such strong interference
that galaxy detection is impossible. We have identified two galaxies in this
image which are marked with arrows in Figure 3. These detections are average in
brightness.  

\subsection{ADBS Source Identification}

\par The two dimensional images (Figure 3) proved to be a good, albeit
time consuming, way to look at these 300,000 spectra. The images allowed us to
easily identify strong interference which showed up in both feeds and often
persisted for several spectra or would appear at the same wavelength a little
later. Without this larger scale view it also would have been easy to
misidentify peaks in the standing wave patterns as sources. Each image was
looked at in tandem with its pair from a different day. With two images, two
feeds, and two polarizations at hand, the reliability of the galaxy
identification was $\sim$58\% even for sources at a signal-to-noise of less than
7 (see \S 5 for full discussion of reliability). From the original ADBS data
we selected 407 sources, 265 of which were later confirmed. 

\par The identification of sources is a delicate balance between the need for
clear-cut selection criteria and the desire for both completeness and
reliability of the sources. Examining the data by-eye proved to be very
effective in finding sources, but it can be difficult to quantify in terms of
the selection criteria. We also used the SExtractor neural network detection
algorithm (Bertin \& Arnouts, 1996) to automate the detection process and make
it easier to quantify. These results are discussed in \S 4. The first tests
quickly proved that the automated procedure was not very good at providing a
reliable data set given only a single image. Because of the noise
characteristics of the data, the program tended to find either myriad
sources or none. The computer algorithm was much more effective when applied to
the data for which there were two independent images. Cross-correlation of the
detection positions dramatically reduced the number of spurious
computer-identified sources. For those images where there was not a second
observation, the human eye proved to be much more effective at selecting
reliable source candidates.

\section{FOLLOW-UP TO THE ADBS}

\subsection{The VLA Follow-Up}

\par The first follow-up observations to the ADBS were made at 21 cm with the
VLA in D-array made in January 1998. With an allotment of 24
hours, it was impossible to follow-up all of the ADBS sources so we opted to
observe as many of the sources with 2 ADBS detections and velocities less than
3000 \kms\ as possible. We took 10 minute snapshots of 99 of the ADBS galaxies.
There were 78 sources observed with velocities of less than 3000 \kms\ in this
data set. Of the 99 sources observed, 90 were detected, proving this to be a
highly reliable data set as expected for the sources with two ADBS detections. 

\par The VLA D-array was used for these observations because the sensitivity of
this configuration allowed us to detect our sources in a reasonable amount of
time and provided good velocity resolution over a large field of view (31\arcm\
HPBW). The synthesized beam-width for this configuration was 44\arcs\ and we
observed at a velocity resolution of 20.5 \kms. With 10 minute snapshot
integrations we were able to achieve an RMS noise of 1.35 mJy/beam at the
primary beam center which is below the detection limits of the ADBS unless the
source was extended over more than about 7 beams. The VLA also provides velocity
field information which is unavailable with a single dish observation. 

\par The VLA data reduction was performed using standard AIPS methods. The
images were created and cleaned using IMAGR with natural weighting. The final
images were then primary beam corrected (PBCOR). The resulting sources were fit
using the Gaussian fitting routine JMFIT which determined the HI centers and
sizes.

\subsection{The Arecibo Follow-up}

\par To refine the positions and fluxes for the remaining ADBS sources, they
were observed with the newly refurbished Arecibo telescope in July 1998, May
1999, and August 1999. To determine the declinations and
fluxes of our sources, we chose to track the telescope at the detection right
ascension and drive the telescope $\pm$7.5\arcm\ from the driftscan declination.
We will refer to this observation method as ``DECscan" mode. The
telescope was driven in declination for a total of 3 minutes and spectra were
dumped every 10 seconds providing 18 spectra separated by  50\arcsper\ The
resulting RMS sensitivity is 1.5 mJy. Figure 4a shows an example of the
resulting data.

\par The flux was determined in each of the individual ``DECscan" spectra and a
Gaussian was fit to the resulting function of total fluxes versus spectrum
position as in Figure 4b. The peak of the derived Gaussian gives the
single-beam flux, and the full width at half-maximum of the fit is a measurement
of the galaxy size. The fit center gives the improved declination position. The
velocity width of the source was determined from a composite spectrum made up of
the sum of the spectra in which the source was detectable.

\par For 10 of the survey sources (020045+2809, 024328+2035, 071028+2307,
084411+2208, \\ \noindent 084504+0932, 113845+2008, 171634+2135, 180506+2308,
183229+2308, 231941+1011) the follow-up Arecibo ``DECscans" confirmed the
detections, but the association with an optical counterpart in the DSS images
was in question.  For these sources, an ``RAscan" was performed, that was
identical to the ``DECscans" except that the telescope was fixed at the
``DECscan" declination and was driven $\pm$7.5\arcm\ in right
ascension to improve the source coordinates. In most cases the tentative optical
counterpart appeared to be confirmed. The source coordinates in Table 2 reflect
the improved ``RAscan" position.

\par A total of 378 sources were re-observed at Arecibo. Out of these 215 were
detected, 31 had been previously detected at the VLA, and 14 sources were
observed during two of the Arecibo observing runs. Multiple observations of the
same sources were used to check the internal consistency of the data set (\S 6).

\section{DATA}

\par Table 1 contains a list of the detection statistics for the Arecibo 
Dual-Beam Survey. We identified 407 possible sources. The reliability of these
sources was 78\% when they were covered twice and 42\% when there was only a
single coverage. Of the 265 sources, 184 have previous optical identifications.

\par Table 2 contains all 265 of the sources that were re-detected on
follow-up. The table contains the RMS of the detection scan or scans which is
typically 3-4 mJy. The rest of the data in the table are the fluxes, line
widths, and velocities from the Arecibo and VLA follow-up observations. 

\par Table 3 is a summary of the ADBS catalog identifications and the optical
data for these sources. This optical information is derived using SExtractor
(Bertin \& Arnouts, 1996) on the DSS images. The red plates of the POSS II
(Second Palomar Optical Sky Survey) images were used when they were available,
but for 48 sources the red POSS I (First Palomar Optical Sky Survey) plates were
used as noted in the table.

\par SExtractor is a routine that first identifies sources in the image, 
extracts them, and determines their physical parameters. The routine
determines the positions of the major and minor axes, and the total number of
pixels exceeding 1$\sigma$. The dimensions, a and b, that SExtractor reports
are the RMS of the intensity along each axis. We determined the ``size" of the
the galaxy from the number of pixels within the 1-$\sigma$ level ellipse: Pixels
= $\pi$ab$\cdot$x$^2$ where x is the factor to scale from RMS to the major or
minor axis size. In most cases x was about 2.5. This sigma level seemed to best
correlate with the visible extent of the galaxies, particularly the ones with
extended low surface brightness regions. 

\par Figure 5 shows a comparison between SExtractor sizes and UGC sizes.
Overall, the SExtractor measurements agree fairly well with the UGC values. We
suspect that most of the scatter is due to inconsistencies in the depth of the
UGC measurement since they were done by-eye on the POSS I, where the resolution
is lower and where the larger spacing between plates allowed for more
variability due to vignetting. In a few of the most extreme cases, the claimed
UGC dimensions appear to extend well beyond any visible emission from the
galaxy. 

\par Table 3 lists only those sources with optical identifications in the 
NASA/IPAC Extragalactic Database (NED\footnote{The Database (NED) is operated by
the Jet Propulsion Laboratory, California Institute of Technology, under
contract with the National Aeronautics and Space Administration.}). Note that 9
of the identified sources come from low surface brightness or dwarf galaxy
surveys like the LSBC (Schombert \et\ 1992), ESDO (Eder \et\ 1989), KK98
(Karachentseva \& Karachentseva 1998), KDG (Karachentseva \et\ 1996). There are
also many larger low surface brightness sources in our lists that were
identified in the UGC catalog (Nilson, 1973), but missed by more traditional
magnitude-limited optical surveys. In addition, there are galaxies at low
Galactic latitude and therefore hidden behind our Galaxy's gas and dust that can
not be detected optically. 30 of the 39 galaxies with visual extinctions $>$ 1.0
are not in any of the optical catalogs.

\section {COMPLETENESS AND RELIABILITY}

\par Completeness is the percentage of galaxies detected in a given volume down
to a defined sensitivity limit. It is important to understand the completeness
of the survey as a function of its detection parameters because this relates
the number of sources found to the number of sources actually present.
Reliability is the percentage of sources initially selected that are
subsequently confirmed. There is a balance between completeness and reliability.
The more sources selected from the original data and the further into the noise
one ``digs," the better the completeness. However, a larger fraction of the
selected sources will be unreliable. Because of the telescope time required for
confirmation, these two must be carefully balanced.

\par For the ADBS data we determine the reliability for both the by-eye and
automated detection methods. The by-eye method proved to be much more reliable.
Overall we confirm 78\% of the possible signals with two original detections
and 42\% for those with only a single detection. The reliability is $>$70\% for
both single and double detections when the signal-to-noise is above 7 but it
falls to 57\% for double detections and 37\% for single detections below this
limit. While the reliability does decrease at lower signal-to-noise values,
especially for sources with only a single detection, the statistics are still
much better than with SExtractor. With SExtractor the reliability was only 12\%
for double detection sources and uncovered no genuine sources not already found
by-eye.

In general, it is not possible to determine the completeness of a survey based
on the detections. It if often assumed that the sensitivity of a survey is
determined by the sources with the lowest signal-to-noise. Schneider {\it et
al.} (1998) showed that ${\cal V}/{\cal V}_{max}$ tests indicate that this
sort of simple sensitivity cutoff does not adequately describe survey
completeness. To quantify the
completeness of the ADBS we added synthetic sources spanning the full range of
galaxy linewidths and with fluxes varying from undetectable to that of bright
galaxies that we expected to detect easily. These sources were given realistic
line profile shapes and were all added to the data prior to the reduction
procedure. 

\par We use the synthetic sources to determine completeness as a function of
signal-to-noise and to compare the results of our by-eye and automated detection
methods. Figure 6 shows the percentage completeness as a function of
signal-to-noise for automated detection (solid line) and by-eye detection
(dotted line). The figure demonstrates the similarity in detection ability of
the two methods, but there is a small improvement using automated
detection. 

We use visual inspection in making our source lists because, though it
slightly decreases the completeness, it greatly improves the reliability. We
note that all of the by-eye detections that were confirmed were also detected by
the automated algorithm. We also ran a test in which we reobserved 25
randomly-selected SExtractor detections. Out of these, 3 were also detected
by-eye and were confirmed whereas none of additional 22 possible signals were
confirmed in follow-up.

Two effects on completeness are apparent in these simulations. First, we confirm
that wide-line sources are not detected at $S/N$ levels quite as low as
narrow-line sources when the noise is assumed to increase as $w^{0.5}$. By
assuming the effective noise grows like $w^{0.75}$, which Schneider \et\ (1998)
adopted based on the ${\cal V}/{\cal V}_{max}$ test, a consistent shape for the
completeness curve is found independent of profile width. Second, with the
statistics of several hundred synthetic sources, it is possible to characterize
the behavior of the roll-off in completeness. It is apparent that occasionally
even high $S/N$ sources are missed and some $<5\sigma$ sources are detected.
These results are consistent for the two detection methods and will be important
for the determination of the mass function in paper II.

Figure 7 is another way of examining the completeness of the sample. The first
panel shows the synthetic sources. The detected sources
are shown as filled gray dots while those we did not detect are shown as open
circles. The completeness line with a slope of 0.75 and a ``S/N" of 5 (using the
formalism of Schneider \et\ 1998) is shown. The line represents a good fit to
the lower limit of the detected points, but it {\bf does not} represent a 100\%
completeness limit. The apparent difference in the completeness for narrower and
wider sources is not a real one. In fact, the completeness above the
``S/N" = 5 line is 78\% for all sources and 80\% for sources with widths
$<$ 300 \kms.  

The second panel in Figure 7 shows fluxes and line widths of confirmed real
sources in the ADBS (gray dots). The data represented here are the confirmed
values for the widths and fluxes adjusted for frequency response and the
offset of the source from beam center. We use these confirmed sources since they
are of higher signal-to-noise and therefore give a better
indication of the true flux. We have eliminated all sources with offsets greater
than 2$\arcm$\ from this plot because of the uncertainties in the measured
flux when sources are beyond the half-power beamwidth. These uncertainties arise
from the effect of small errors in the pointing and a lack of complete
information about the HI extent.

Figure 8 shows the percentage completeness for the synthetic sources as a
function of velocity. Notice the spikes in incompleteness near either end and in
the center of the velocity range. These correspond to the ends of the bandpass
segments. The baseline fitting at the ends of the segments is particularly
prone to problems with source subtraction because there is not baseline on
either side of the feature to distinguish a source from a baseline ``wiggle."
There is the additional problem for real sources in these regions that the
sensitivity is falling off rapidly at the bandpass edges. If we eliminate
sources with velocities $<$ 300 \kms\ (to deal with confusion with the Milky Way
too) and between 3100 and 3600 \kms\ the data are 82\% complete above our
``5$\sigma$ completeness" limit.

\par The ideal check of completeness is to have a much deeper survey over the
same area of sky, with the same instrument, with which to compare your results.
Zwaan \et\ (1997) provide us with just such a comparison sample for a small
section of our survey. Their sample consists of $\sim$ 10 hour driftscans at 2
declinations covered repeatedly to achieve about 5 times the sensitivity of our
survey (RMS noise = 0.75 mJy in 16 \kms\ channels). One of
these strips, at $\sim$23\degre, overlaps one of ours within 20\arcsper\ Because
their data are much more sensitive, we can test our completeness although small
number statistics make it impossible to use these results for a detailed study
of sensitivity roll-off.

The triangles in Figure 7 are the Zwaan \et\ sources; the filled triangles are
the sources we detect, the open triangles are the ones we miss. We have detected
all of the Zwaan \et\ sources above the ``S/N" = 5 line. There are two sources
that we pick up as a single source. They are plotted with the rest of the
sources and one falls with the detections, so we probably would have detected it
on its own, while we probably would have missed the fainter one. 

\section{PROPERTIES OF THE SAMPLE}

\subsection{Position Determination}

\par Figures 9a and 9b provide an evaluation of the positional accuracy of
the ADBS, VLA, and ``DECscan" measurements. We compare positions to the optical
centers when the optical association was clear or to the VLA positions when
the association was not. The optical centers were determined from the DSS
images. Figure 9a shows the uncertainty in the RA measurements. The open
histogram is the offset between the detection positions and either the VLA or
the optical positions. These offsets have a standard deviation of 57\arcs, about
half of the 105\arcs\ separation between spectra in the detection survey. The
shaded histogram is the offset between the VLA positions and the optical
positions, these offsets have a standard deviation of 31\arcsper\ This standard
deviation is larger than expected from the measurement uncertainties of the
optical and VLA positions and may be partially due to asymmetries in the HI
distribution with respect to the optical emission. 

Figure 9b shows the uncertainty in the original declination positions also
using either the VLA or optical counterpart positions. The open histogram is the
offset between the detection declinations and either the VLA or the optical
positions. The standard deviation in this distribution is 2.1\arcm\ which is
about two-thirds of the width of the main beam, but there are several sources
that are detected in the sidelobes up to 7 \arcm\ from the VLA or optical
centers. The sources with the biggest offsets are all large galaxies that we
catch the edge of. The shaded histogram shows the offsets between the Arecibo
``DECscan" positions and the VLA or optical positions. The standard deviation in
these offsets is 29\arcs, similar to the VLA positional uncertainties (which are
not shown, but are the same as for the right ascension). Figures 9a and 9b
demonstrate that the final positions have uncertainties of $\sim$30\arcs\ for
the VLA and ``DECscan" coordinates, but 2.1\arcm\ in right ascensions for
sources that were not re-observed at the VLA.

\par Figure 10 is an example of why our detection method does not always
determine the best position for an object. The galaxy in this figure was
detected in a slice that passed to its north. Because of the galaxy's position
angle, the detection position was too far to the east. The ``DECscan" improved
the measurement, but it was still affected by the galaxy's position angle. For
the larger galaxies, these positioning problems can cause substantial errors. 

\subsection{Flux Determination}

\par The flux measurement for a source will be underestimated if it is offset
from the center of the beam. As we have shown in \S 6.1, our initial ADBS
positions are up to 7$\arcm$\ from the galaxy center. Our follow-up observations
have much higher S/N than the detection scans because the integration time is
longer and because the sources are better centered. We have used several methods
to evaluate the accuracy and the internal consistency of the follow-up flux
measurements. The crosses in Figure 11a show the fluxes of Arecibo sources
measured in both July 1998 and May 1999, showing the consistency of the
measurements on 2 independent dates. The filled circles demonstrate the
consistency between an on-off and the original ``DECscan" value. To further test
the accuracy of the flux determinations, we made ``DECscan" observations of
several UGC galaxies for which there are high quality, single-beam HI data in
the literature. Figure 11b shows the relationship between our flux determination
and the literature values for these UGC galaxies. Overall, our fluxes agree well
with the literature values although a few of the literature measurements are
somewhat high. We reobserved three of the most discrepant sources with on-off
measurements and found values in better agreement with our ``DECscan" fluxes
suggesting that our fluxes are not underestimated.

\subsection {Spatial Distribution}

\par The ADBS survey covers velocities from --654 $<$ v $<$ 7977 \kms. Figure 12
shows the distribution of HI sources with redshift. No sources were found
at negative velocities in the sample, and only three sources were detected with
v $>$ 7500 \kms. At low velocities, we avoided selecting high velocity
clouds that were highly extended in right ascension. We found a few candidate
signals in this velocity range, but none were confirmed. Figure 12 shows that
the distribution of galaxies in velocity space is fairly flat; this is probably
due to surveying many different environments over a large area on the sky which
tends to smear the signatures of large scale structure.

\par The redshift-position distributions of the galaxies are shown in Figures
13 a and b. We have broken the sample into a low declination region and a high
declination region and plot the redshift versus RA. The stars in the plot
represent ADBS sources that are overlayed on gray dots representing the RC3 (de
Vaucouleurs \et\ 1991) sources. It is evident from these figures that, at least
qualitatively, the ADBS sources follow the structures defined by the RC3
galaxies. This is consistent with the findings of Spitzak \& Schneider (1998)
and Zwaan \et\ (1997) for their ``blind" HI samples.

\subsection {Optical Counterparts}

\par The DSS provides us with a look at the optical nature of our survey
sources. The POSS I and POSS II sizes are comparable, although the relationship
has a lot of scatter. Sources with only POSS I images are footnoted in Table 3.
We find a wide range of optical properties for the galaxies detected in this
survey. The galaxy types range from average spirals with angular diameters from
15.5\arcm\ (NGC 3628) to compact galaxies only a few arcseconds across to
galaxies that appear to be extremely faint patches of nebulosity. Figure 14
shows a histogram of the angular sizes of the survey sources. Some sources, like
those with just a bit of wispy nebulosity and low Galactic latitude objects,
were not measurable optically so they are not included. While the mean size is
$\sim$100\arcs, the majority of the sources are smaller than 70$\arcsper$\ 

Table 4 lists all of the sources for which no optical counterpart was firmly
identified. There are 11 galaxies with A$_v$ $>$ 2 mag that are not discernible
on the POSS plates. All but one of these are in the Zone of Avoidance (ZOA)
hidden by the gas, dust, and stars of our own galaxy. The one exception is an
unusual case of a source at -34\degre\ latitude which is behind a very small
dense clump of gas identified in the Schlegel \et\ survey (1998) to have an
A$_v$ $>$ 6 mag. There are also 11 sources in the list which are not in fields
of high extinction where an optical identification is not obvious. Figure 15
shows the DSS images for 4 of these extremely faint or optically ambivalent, in
the case of 142335+2131, galaxies. We note that these 11 sources do not include
several other very low surface brightness counterparts where we feel the match
is unambiguous.

\subsection {Masses}

\par A thorough analysis of the HI and dynamic masses for the ADBS galaxies is
left for paper II, but Figure 16 shows the distribution of HI masses to
demonstrate the range of source types we detect. The masses are determined using
the standard conversion from flux: 

\begin{equation}
 M_{HI} = 2.36\times10^5\cdot D^2\cdot \int S{\it dv} 
\end{equation} 

The distance used to determine the HI mass is corrected by the potential flow
algorithm, POTENT (Bertschinger \et\ 1990), or V$_0$ for galaxies where the
correction was undefined. The sample contains sources ranging in mass from
2.3$\times$10$^7$ to 1.5$\times$10$^{10}$ M\solar. There are 7 galaxies with
masses $<$ 10$^8$ M$\solar$. This is relative to 4 sources in Spitzak \&
Schneider (1998) and 2 for Zwaan \et\ (1997; assuming H$_0$ = 75 \kms\
Mpc$^{-1}$). 

\section{SUMMARY}

\par The size of HI identified samples is growing with the contributions of
Zwaan \et\ (1997), Spitzak \& Schneider (1998), HIPASS (Kilborn \et\ 1999) and
this survey. This driftscan survey has contributed a sample of 265 sources to
the endeavor, 81 of which were not previously cataloged. All of these sources
have been confirmed and have improved positions and fluxes based on follow-up
VLA and upgraded-Arecibo measurements.

\par More important than numbers detected first optically or in HI is a thorough
understanding of completeness. To make an accurate assessment of the
completeness of our sample we have used synthetic sources which were
incorporated into the data prior to the data reduction. We are not 100\%
complete at any signal-to-noise level, but we have developed a clear picture of
the sensitivity roll-off.

\par We have evaluated the positional accuracy of the survey and find that the
original detection survey has a positional accuracy of $\sim$1$\arcm$ in RA and
$\sim$2$\arcm$ in DEC. The follow-up VLA and Arecibo scanning measurements
improve the positions to an accuracy of $\sim$30$\arcsper$ These
positions have allowed us to measure the probable optical counterparts in all
but 22 cases, half of which are in the Galactic Plane.

Using the DSS we have examined the optical nature of the HI sources detected in
this survey and find a wide range of optical properties for the galaxies. There
are Messier objects, there are faint wisps of nebulosity, and there are 11
galaxies completely enshrouded in the dust of the Milky Way so as to be
indiscernible on the DSS images. A second set of 11 sources have no clear
optical counterpart or wisps of nebulosity that nay be associated.

\par The flux measurements for these galaxies are repeatable and are in good
agreement with values from the literature. The detection fluxes, and in a few
cases the confirmation fluxes, can be affected by the offset of the galaxy from
beam center particularly for large galaxies. There does not, however, appear to
be any discrepancy between fluxes determined from our special scanning technique
and standard on/off measurements.

\par From a qualitative look at the spatial distribution of our sample of
galaxies relative to the distribution of sources in the RC3 (de Vaucouleurs
\et\ 1991), it is evident that the HI-selected sources follow the structures
defined by the RC3 sample. This is consistent with the findings of Spitzak \&
Schneider (1998) and Zwaan \et\ (1997) for their ``blind" HI surveys.

The masses of galaxies in this survey range from 2$\times$10$^7$ M$\solar$ to
1$\times$10$^{10}$ M$\solar$. We find 7 galaxies with masses $<$ 10$^8$
M$\solar$ -- almost as many as all of the previous ``blind" HI surveys combined.
This data set, in conjunction with the samples of Spitzak \& Schneider (1998)
and Zwaan \et\ (1997), should help constrain the faint end of the HI mass
function. The larger number statistics and use of synthetic sources to quantify
our detection limits should prove extremely powerful in applying this data to
determining the HI mass function. Paper II will investigate the mass function as
determined by these galaxies in detail.

\acknowledgements

We would like to thank the staffs at Arecibo and the VLA for their assistance 
with the observations. We would also like to thank Frank Briggs and
Ertugrul Sorar for helpful discussions about this work. This work was supported
in part by NSF Presidential Young Investigator award AST-9158096, and 2 Sigma-Xi
Grants-in-Aid of Research.

The Digitized Sky Surveys were produced at the Space Telescope Science Institute
under U.S. Government grant NAG W-2166. The images of these surveys are based on
photographic data obtained using the Oschin Schmidt Telescope on Palomar
Mountain and the UK Schmidt Telescope. The plates were processed into the
present compressed digital form with the permission of these institutions. 

The National Geographic Society - Palomar Observatory Sky Atlas (POSS-I) was
made by the California Institute of Technology with grants from the National
Geographic Society. 

The Second Palomar Observatory Sky Survey (POSS-II) was made by the California
Institute of Technology with funds from the National Science Foundation, the
National Geographic Society, the Sloan Foundation, the Samuel Oschin Foundation,
and the Eastman Kodak Corporation. 

The Oschin Schmidt Telescope is operated by the California Institute of
Technology and Palomar Observatory. 

The UK Schmidt Telescope was operated by the Royal Observatory Edinburgh, with
funding from the UK Science and Engineering Research Council (later the UK
Particle Physics and Astronomy Research Council), until 1988 June, and
thereafter by the Anglo-Australian Observatory. The blue plates of the southern
Sky Atlas and its Equatorial Extension (together known as the SERC-J), as well
as the Equatorial Red (ER), and the Second Epoch [red] Survey (SES) were all
taken with the UK Schmidt.

\figcaption{The sky coverage of the ADBS survey. The thick lines show the areas    
where  the sky was covered twice, the thin lines show where it was covered only 
once. The narrow lines on either side of each coverage strip show the positions 
of the  sidelobes.}                                                             

\figcaption{A cross-section of the effective convolved sensitivity of the Arecibo  
beam as is moves across the sky during a driftscan.}                            

\figcaption{An example of one of the FITS images created by stacking 300
spectra. The x-axis represents the velocity, or distance, dimension
while the y-axis represents RA. The image is split vertically into the
two feeds which are actually spectra of different regions of the sky
observed simultaneously. Along the horizontal, there is a pattern in
the center of the image which is where additional information about
the spectra has been stored. To the left of that, the data displayed
are the average of the two polarizations, while to the right the image is
the absolute value of the polarization difference.}

\figcaption{(a) Eight spectra (out of 18 total) in which a source
was detected in a ``DECscan." The y-axis is a measurement of the flux of the
source but each spectrum is offset for display purposes. (b) The flux in each of
the 18 spectra. This function is fit with a  gaussian to determine the flux and
declination of the source, as well as a rough measure of the HI extent.}

\figcaption{Comparison of SExtractor sizes and UGC sizes. The most errant points
are generally low surface brightness galaxies for which SExtractor terminates
too early, or for which the extent of the galaxy on the POSS I images, upon
which the UGC was based, was unclear.}

\figcaption{Completeness of the ADBS detection as determined by-eye (dotted line)
and by computer algorithm (solid line) using synthetic sources inserted in the
data.}

\figcaption{The left-hand panel shows the relationship between flux and line       
width for the synthetic sources inserted into our data. The gray, filled circles
are the sources we detect. The open circles are the sources we miss. The solid  
line represents the completeness limit of the survey as defined in Schneider    
\et\ (1998) with a S/N constant of 5 and a slope of 0.75. The figure shows that 
even above the ``completeness" limit we miss 20\% of the sources. The second    
panel shows the same thing for real data. The filled dots are the sources       
detected in our survey. The filled triangles are the sources from Zwaan \et\    
(1997) that we also detect. The open triangles are the Zwaan \et\ sources that  
we miss. The gray triangles are 2 galaxies detected as one in both our survey   
and Zwaan \et. We probably would detect the brighter source by itself, but not  
the fainter one.}                                                               

\figcaption{A histogram showing the percentage of sources above our
``completeness" limit that are missed in each velocity bin.}

\figcaption {(a) The open histogram in shows the difference between               
the RA detection positions and the VLA or optical centers. The filled histogram
shows the difference in RA between the VLA and the optical centers. (b) The open
histogram shows DEC difference between the detection positions and the VLA or  
optical centers. The filled histogram shows the difference between the         
``DECscan" positions and the VLA or optical centers.}                          

\figcaption{A demonstration of what can happen when determining the position of a 
galaxy which is not aligned with a cardinal direction. The original driftscan  
determined the incorrect right ascension and our ``DECscans" do not provide any
additional information about the East-West location of the galaxy.}            

\figcaption{(a) A comparison between the Arecibo fluxes measured during each of   
the two Arecibo follow up runs is shown by the crosses. The filled circles show
a comparison between the ``DECscan" flux and an on-off measurement of the same 
source.(b) A comparison between the flux values measured in this survey and the
literature values for UGC sources that are cataloged and have high quality     
Arecibo flux measurements in the literature. The symbols in the plot refere to 
papers as follows: $\times$ Haynes \& Giovanelli 1984, $\circ$ Schneider \et\  
1990, $\triangle$ Bicay \& Giovanelli 1986, $\ast$ Giovanelli \et\ 1986,       
$\bullet$ Freudling \et\ 1988, $\bull$ Lewis 1987, and $\Box$                  
Giovanelli \& Haynes 1989.}                                                    

\figcaption{Histogram of the heliocentric velocities for the galaxies detected in 
the ADBS survey.}                                                              

\figcaption{The redshift distribution of the ADBS galaxies, black stars,          
relative to RC3 sources, gray dots, out to cz = 8000 \kms. (a) The declination 
range 8\degre\ to 18\degre. (b) The declination range 18\degre\ to 28\degre. } 

\figcaption{Histogram of the SExtractor determined optical diameters from the DSS 
images.}                                                                       

\figcaption{The DSS images of 4 galaxies with faint, questionable optical         
counterparts.}                                                                 

\figcaption{Histogram of the HI masses of the ADBS sources.}

\begin{deluxetable}{rccc}
\tablewidth{5in}
\tablecaption{Detection Statistics}
\tablehead{
\colhead{} & \colhead{N} & \colhead{1 observation} &
\colhead{2 observations}}
\startdata
Total Linear Coverage & \nodata & 132.3 hrs & 399.3 hrs \\
& & & \\
Suspected Sources & 407 & 146 & 261 \\
& & & \\
Not Confirmed & 142 & 85 & 57 \\
Confirmed & 265 & 61 & 204 \\
& & & \\
Cataloged & 184 & 50 & 134 \\
Not Cataloged & 81 & 11 & 70 \\
\enddata
\end{deluxetable}

\clearpage
\begin{deluxetable}{l@{ }c@{ }r@{ }l@{ }cc@{ }c@{ }c@{ }c@{ }cc@{ }r@{ }r@{ }r@{ }r@{ }r@{ }r@{ }}
\tabletypesize{\scriptsize}
\rotate
\tablecaption{Galaxy Detection Data}
\tablewidth{8.5in}
\tablehead{
\colhead{Name} & \colhead{N$_{obs}$ \footnotemark[1]} & \multicolumn{3}{c}{RA} &
\multicolumn{3}{c}{DEC} & \colhead{RMS \footnotemark[2]} &
\colhead{RMS \footnotemark[2]} & \colhead{A Flx \footnotemark[3]} &
\colhead{A $\Delta$V$_{20}$ \footnotemark[4]} &\colhead{A $\Delta$V$_{50}$ \footnotemark[5]} &
\colhead{A Vel \footnotemark[6]} & \colhead{V Flx \footnotemark[7]} &
\colhead{V $\Delta$V$_{50}$ \footnotemark[8]} & \colhead{V Vel} \footnotemark[9] \\
\colhead{} & \colhead{} & \multicolumn{3}{c}{(2000)} & \multicolumn{3}{c}{(2000)} &
\colhead{mJy} & \colhead{Jy \kms} & \colhead{\kms} & \colhead{\kms} &
\colhead{\kms} & \colhead{Jy \kms} & \colhead{\kms} & \colhead{\kms} & \colhead{\kms}}
\startdata
000330+2312 & 2 & 00 & 03 & 30.5  & 23 & 12 & 03.6 & 3.41 &   3.39  &    4.663 &  363.355 & 338.732 & 7313.468 & \nodata & \nodata &  \nodata \\
000407+2234 & 2 & 00 & 04 & 07.0  & 22 & 34 & 55.2 & 3.49 &   3.51  &    2.858 &  226.510 & 219.58 & 4475.981 & \nodata & \nodata &  \nodata \\
000623+2347 & 2 & 00 & 06 & 23.0  & 23 & 47 & 31.2 & 3.57 &   3.62  &    2.618 &  337.622 & 296.021 & 4682.532 & \nodata & \nodata &  \nodata \\
000900+2348 & 2 & 00 & 09 & 00.5  & 23 & 48 & 57.6 & 3.53 &   3.67  &    2.233 &  279.771 & 237.790 & 4492.880 & \nodata & \nodata &  \nodata \\
001622+2237 & 2 & 00 & 16 & 22.6  & 22 & 37 & 04.8 & 3.54 &   3.51  &    3.416 &  325.708 & 249.79 & 5784.375 &   3.322 & 252.400 & 5819.007 \\
002249+2310 & 2 & 00 & 22 & 49.2  & 23 & 10 & 19.2 & 3.42 &   3.45  &    1.179 &   48.448 &  34.961 & 4525.693 & \nodata & \nodata &  \nodata \\
002526+2136 & 2 & 00 & 25 & 26.4  & 21 & 36 & 14.4 & 3.50 &   3.56  &    1.582 &  142.456 & 116.464 & 4631.979 &   3.025 & 169.800 & 4592.007 \\
003426+2436 & 2 & 00 & 34 & 26.6  & 24 & 36 & 03.6 & 3.44 &   3.52  &    3.810 &  164.503 & 146.775 & 5341.837 & \nodata & \nodata &  \nodata \\
003751+0838 & 2 & 00 & 37 & 51.4  & 08 & 38 & 31.2 & 3.64 &   3.68  &    5.826 &  394.050 & 355.016 & 5278.171 & \nodata & \nodata &  \nodata \\
003811+2523 & 2 & 00 & 38 & 11.0  & 25 & 23 & 45.6 & 3.46 &   3.43  &    1.372 &  128.032 &  91.848 & 5201.868 &   2.261 & 102.800 & 5229.003 \\
004649+2134 & 2 & 00 & 46 & 49.0  & 21 & 34 & 58.8 & 3.40 &   3.52  &    1.112 &   59.219 &  92.618 & 5170.682 & \nodata & \nodata &  \nodata \\
011440+2708 & 2 & 01 & 14 & 40.3  & 27 & 08 & 06.0 & 3.47 &   3.49  &    3.261 &  144.113 & 109.525 & 3620.157 & \nodata & \nodata &  \nodata \\
014206+1235 & 1 & 01 & 42 & 06.7  & 12 & 35 & 60.0 & 3.58 &  \nodata  &   18.494 &  339.638 & 305.975 & 3065.943 & \nodata & \nodata &  \nodata \\
014246+1309 & 1 & 01 & 42 & 46.8  & 13 & 09 & 18.0 & 3.65 &  \nodata  &    8.125 &  152.775 & 129.193 &  807.862 & \nodata & \nodata &  \nodata \\
014527+2531 & 2 & 01 & 45 & 27.6  & 25 & 31 & 19.2 & 3.54 &   3.57  &    7.417 &  110.014 &  91.536 & 3845.949 & \nodata & \nodata &  \nodata \\
014729+2719 & 2 & 01 & 47 & 29.5  & 27 & 19 & 58.8 & 3.56 &   3.52  &   39.051 &  129.126 &  94.210 &  334.825 &  62.860 & 104.900 &  375.707 \\
014847+1034 & 2 & 01 & 48 & 47.0  & 10 & 34 & 08.4 & 3.64 &   4.40  &    1.512 &  241.366 & 134.979 & 5245.624 &   1.587 & 146.500 & 5283.000 \\
015011+2309 & 2 & 01 & 50 & 11.8  & 23 & 09 & 21.6 & 3.53 &   3.41  &    1.485 &  131.067 &  90.586 & 2863.507 &   2.934 &  87.690 & 2886.000 \\
015105+1235 & 1 & 01 & 51 & 05.3  & 12 & 35 & 16.8 & 3.65 &  \nodata  &    2.144 &  273.356 & 183.402 & 3280.678 & \nodata & \nodata &  \nodata \\
015434+2312 & 2 & 01 & 54 & 34.3  & 23 & 12 & 14.4 & 3.51 &   3.49  &  \nodata &  \nodata & \nodata &  \nodata &   3.482 &  95.810 & 5045.000 \\
015443+1033 & 1 & 01 & 54 & 43.2  & 10 & 33 & 57.6 & 3.54 &  \nodata  &    1.781 &   49.790 &  32.086 & 6086.814 & \nodata & \nodata &  \nodata \\
015906+2523 & 2 & 01 & 59 & 06.2  & 25 & 23 & 09.6 & 3.72 &   3.50  &    3.205 &  247.220 & 231.736 & 5066.079 & \nodata & \nodata &  \nodata \\
020022+2434 & 2 & 02 & 00 & 22.8  & 24 & 34 & 48.0 & 3.53 &   3.44  &    4.055 &  256.081 & 228.280 & 5134.388 & \nodata & \nodata &  \nodata \\
020045+2809 & 2 & 02 & 00 & 45.2  & 28 & 09 & 57.6 & 3.43 &   3.46  &    2.069 & 66.911 &  48.819 & 5343.904 & \nodata & \nodata &  \nodata \\
020148+2632 & 2 & 02 & 01 & 48.7  & 26 & 32 & 49.2 & 3.68 &   3.82  &    4.134 & 356.984 & 266.742 & 5021.338 & \nodata & \nodata &  \nodata \\
020320+1837 & 2 & 02 & 03 & 20.2  & 18 & 37 & 48.0 & 5.80 &   3.51  &  \nodata & \nodata & \nodata &  \nodata &  10.250 &  86.490 & 2395.007 \\
020320+2345 & 2 & 02 & 03 & 20.9  & 23 & 45 & 39.6 & 3.49 &   3.78  &    1.749 & 230.609 & 202.268 & 2792.723 &   3.095 & 115.600 & 2855.007 \\
020405+2412  & 2 &  02 & 04 & 05.0  & 24 & 12 & 32.4 &   3.57 &   3.50  &    3.110 &   72.172 &  36.86 &  609.264 &   3.243 &  47.200 &  641.707 \\
020918+2534  & 2 &  02 & 09 & 18.0  & 25 & 34 & 15.6 &   3.39 &   3.65  &    4.081 &  406.151 & 337.973 & 4942.253 & \nodata & \nodata &  \nodata \\
022807+1935  & 2 &  02 & 28 & 07.7  & 19 & 35 & 34.8 &   3.47 &   3.55  &   12.519 &  415.978 & 394.993 & 4205.212 & \nodata & \nodata &  \nodata \\
022859+2808  & 2 &  02 & 28 & 59.0  & 28 & 08 & 42.0 &   3.51 &   3.49  &  \nodata &  \nodata & \nodata &  \nodata &   6.287 & 101.600 & 1036.000 \\
\enddata
\end{deluxetable}

\begin{deluxetable}{rrrrrrrr}
\voffset -1.4in
\tablewidth{6.1in}
\tablecaption{Optical Source Information \label{tab:optical}}
\tablehead{
\colhead{ID} & \colhead{Opt. cat. name} & \colhead{l} & \colhead{b} &
\colhead{A$_v$} & \colhead{Diam \tablenotemark{1}} & \colhead{Ell \tablenotemark
{1}}
& \colhead{PA \tablenotemark{1}}}
\startdata

000330+2312  &    UGC               14 & 108.8479 & -38.3574 & 0.406  &  102.675 &    0.392 &  -64.200  \\
000407+2234  &    UGC               24 & 108.8361 & -38.9910 & 0.287  &   60.010 &    0.408 &  -41.800  \\
000623+2347  &    KUG         0003+235 & 109.8355 & -37.9313 & 0.710  &   56.604 &    0.631 &  -50.000  \\
000900+2348  &    NGC                9 & 110.5862 & -38.0365 & 0.403  &   78.439 &    0.577 &   69.300  \\
001622+2237  & \nodata                 & 112.4226 & -39.5372 & 0.379  &   42.926 &    0.583 &    9.100  \\
002249+2310  & \nodata                 & 114.4323 & -39.2270 & 0.253  &   30.229 &    0.285 &   79.400  \\
002526+2136  & \nodata                 & 114.9329 & -40.8636 & 0.293  &   23.963 &    0.339 &   53.300  \\
003426+2436  &    UGC              337 & 118.0211 & -38.1087 & 0.135  &   50.813 &    0.191 &   27.900  \\
003751+0838  &    NGC              180 & 117.2043 & -54.0786 & 0.240  &  126.776 &    0.320 &   81.400  \\
003811+2523  & \nodata                 & 119.1644 & -37.3788 & 0.130  &   38.700 &    0.203 &   19.700  \\
004649+2134  &   ESDO          F540-04 & 121.5026 & -41.2759 & 0.134  &   38.463 &    0.140 &  -78.700  \\
011440+2708  &   ESDO          F475-04 & 129.2796 & -35.4502 & 0.247  &   53.476 &    0.114 &   55.400  \\
014206+1235  &    NGC              658 & 141.7429 & -48.4122 & 0.194  &  175.100 &    0.523 &  -64.300  \\
014246+1309  & UGC 1200 \tablenotemark{3} & 141.7314 & -47.8338 & 0.200 & 124.038 & \nodata & \nodata  \\
014527+2531  &    UGC             1230 & 137.9862 & -35.7655 & 0.477  &   60.156 &    0.238 &   22.500  \\
014729+2719  &     IC             1727 & 137.9526 & -33.8985 & 0.340  &  333.036 &    0.691 &   22.450  \\
014847+1034  &    UGC             1268 & 145.0964 & -49.8188 & 0.517  &   86.773 &    0.673 &  -43.800  \\
015011+2309  &    UGC             1294 & 140.0824 & -37.7510 & 0.469  &   45.526 &    0.308 &  -33.600  \\
015105+1235  &    UGC             1314 & 144.8569 & -47.7283 & 0.303  &   80.921 &    0.587 &   72.200  \\
015434+2312  &   LSBC          F477-01 & 141.2767 & -37.4084 & 0.469  &   34.931 &    0.221 &   84.600  \\
015443+1033  &   ESDO          F685-07 & 147.1993 & -49.3008 & 0.491  &   51.559 &    0.237 &  -38.400  \\
015906+2523  &    UGC             1462 & 141.6537 & -35.0137 & 0.528  &   83.242 &    0.432 &  -13.000  \\
020022+2434  &     IC             1764 & 142.3063 & -35.6845 & 0.505  &   88.731 &    0.223 &  -71.300  \\
020044+2809  & \nodata                 & 141.0223 & -32.2591 & 0.263  & \nodata  & \nodata  & \nodata  \\
020148+2632  &    UGC             1510 & 141.9026 & -33.7190 & 0.316  &   60.015 &    0.475 &  -40.100  \\
020320+1837  &    UGC             1546 & 145.7422 & -41.0354 & 0.325  &   70.216 &    0.057 &  -52.900  \\
020320+2345  &    UGC             1538 & 143.4316 & -36.2317 & 0.381  &   43.730 &    0.277 &  -25.600  \\
020405+2412  &    UGC             1561 & 143.4377 & -35.7533 & 0.362  &  118.698 &    0.294 &   10.007  \\
020918+2534  &    UGC             1648 & 144.2080 & -34.0635 & 0.325  &   90.083 &    0.454 &  -17.200  \\
022807+1935  & NGC 935 \tablenotemark{3} & 152.1288 & -37.7363 & 0.823 & 140.072 & \nodata  & \nodata   \\
022859+2808  &    UGC             1958 & 147.7849 & -29.9767 & 0.576  &   69.318 &    0.657 &  -72.400  \\
\enddata
\end{deluxetable}

\begin{deluxetable}{rrrll}
\tablenum{4}
\tablewidth{6.8in}
\tablecaption{Sources Without an Identified Optical Counterpart}
\tablehead{
\colhead{ID} & \colhead{b} & \colhead{A$_v$} & \colhead{Plate} & \colhead{Notes}}
\startdata
020044+2809 & -32.2591 & 0.263 & POSS II & May be associated with large nearby galaxy \\
024328+2035 & -35.1039 & 0.508 & POSS II & Faint nebulosity behind bright star \\
025537+1938 & -34.3980 & 6.438 & POSS II & High A$_v$ even though it is 34\degre\ out of the Plane \\ 
044753+2346 & -13.5452 & 3.249 & POSS I & High A$_v$ \\
053017+2233 &  -6.3132 & 1.768 & POSS II & Poor image due to bright star, but nothing obvious \\
055452+1509 &  -5.1948 & 1.379 & POSS I & High stellar density, but nothing galaxy-like \\
055517+2526 &   0.0552 & 4.287 & POSS II & Galactic Plane \\
060719+1936 &  -0.4254 & 6.776 & POSS I & Galactic Plane \\
060733+0932 &  -5.2488 & 2.210 & POSS I & Galactic Plane \\
061543+1110 &  -2.7069 & 3.419 & POSS II & Galactic Plane \\
061729+2807 &   5.6839 & 2.313 & POSS II & Galactic Plane \\
062054+2008 &   2.6385 & 3.416 & POSS II & Galactic Plane \\
063549+1107 &   1.6357 & 6.332 & POSS I & Galactic Plane \\
063603+1109 &   1.6983 & 7.149 & POSS I & Galactic Plane \\
084411+2208 &  34.1423 & 0.166 & POSS II & 3 faint sources, unclear ID \\
084504+0932 &  29.5464 & 0.265 & POSS II & Lots of small fuzzy things in field, unclear ID \\
090552+2520 &  39.8224 & 0.160 & POSS II & Many fuzzy sources, unclear ID \\
142335+2131 &  68.3076 & 0.123 & POSS II & 2 faint sources straddling detection position \\
183229+2308 &  14.2863 & 0.513 & POSS II & Possibly faint nebulosity behind bright star \\
184335+2007 &  10.6904 & 1.141 & POSS II & High stellar density, but nothing galaxy-like \\
184829+1835 &   8.9901 & 1.379 & POSS II & High stellar density, but nothing galaxy-like \\
192728+2012 &   1.5309 & 14.801 & POSS I & Galactic Plane \\
\enddata 
\end{deluxetable}                                                               
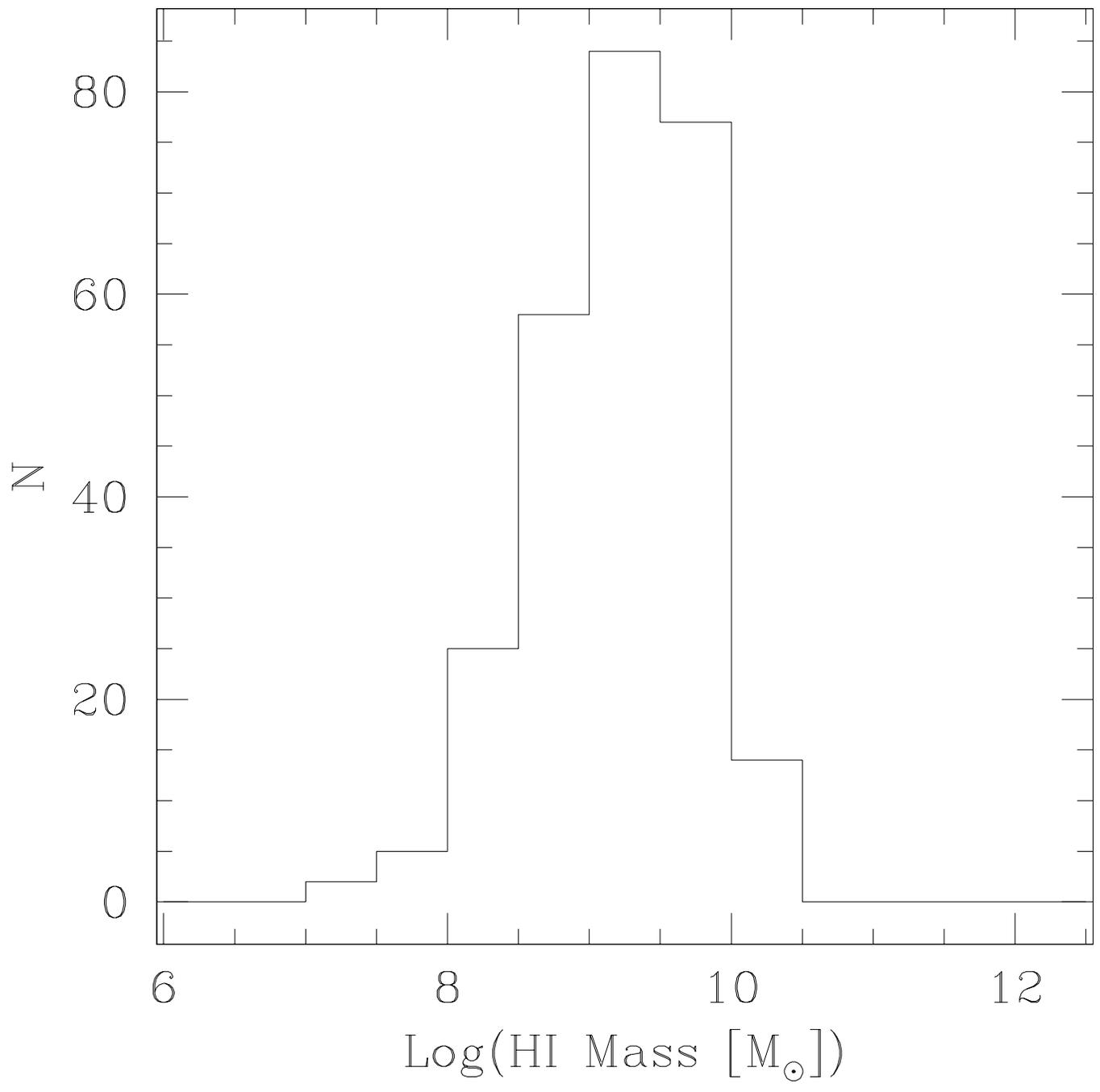
\end{document}